\documentclass[twocolumn,showpacs,preprintnumbers,amsmath,amssymb]{revtex4}
\usepackage[utf8]{inputenc}
\usepackage[english]{babel}
\usepackage{graphicx}
\usepackage{dcolumn}
\usepackage{bm}

\begin{document}

\title{Zahn’s Theory of Dynamical Tides and Its Application to Stars}
\author{S.V. Chernov}
\affiliation{Astrospace Center, Lebedev Physical Institute, Russian Academy of Sciences, Profsoyuznaya ul. 84/32,
Moscow, 117997 Russia}
\email{chernov@lpi.ru}

\begin{abstract}
Zahn’s theory of dynamical tides is analyzed critically. We compare the results of this theory
with our numerical calculations for stars with a convective core and a radiative envelope and with masses of one and a half and two solar masses. We show that for a binary system consisting of stars of one and a half or two solar masses and a point object with a mass equal to the solar mass and with an orbital period of one day under the assumption of a dense spectrum and moderately rapid dissipation, the evolution time scales of the semimajor axis will be shorter than those in Zahn’s theory by several orders of magnitude
\end{abstract}

\maketitle

\section{Introduction}

Zahn's asymptotic theory of dynamical tides is
one of the first theories of dynamical tides developed
in the 1970s for stars with a convective core and
a radiative envelope (Zahn 1970). This theory has
both advantages and disadvantages. One advantage
of this theory is that although it is rather cumbersome,
its corollaries are very simple. It turns out
that the intensity of tidal interactions in this theory is
determined by the dependence of the so-called overlap
integral Q (or the resonance coefficient defined
in Zahn's papers, 1970, and proportional to Q; for
their definition see below) on the stellar normal mode
frequency, with this dependence being a power law
$Q\sim\omega^{17/6}$. According to this theory, the power law
dependence does not evolve with time and remains
valid for any stars satisfying some conditions
to be discussed below. In reality, as our numerical
calculations show, it is quite difficult to obtain this
dependence (Chernov et al. 2013), and many of the
assumptions that were made in deriving this relation
hold only for some very short interval of the stellar
lifetime. Nevertheless, owing to its simple result, this
theory is used in a large number of papers to explain
various observed phenomena (Zahn 1977; Claret and
Cunha 1997; Goodman and Dicksun 1998).

For close binary systems those stellar eigenfrequencies
that resonantly interact with the external
gravitational potential will play a major role. For
a point object revolving around a star with an orbital
period of the order of a few days, and precisely
such orbital periods are believed to be able to lead to
significant changes in the orbital parameters of the
binary system, the resonant eigenfrequencies will fall
within the frequency range $0.1 < \sigma < 1$, where the
dimensionless frequency $\sigma = \omega/\sqrt{GM/R^3}$. This frequency
range refers to the so-called g-modes (Cowling
1941). It is the g-modes that play a crucial role
in the theory of dynamical tides. The g-mode spectrum
is assumed to be sufficiently dense. This means
that the distance between neighboring frequencies is
much smaller than the frequency itself, $\omega_j-\omega_{j+1}<<\omega_j$.
For main-sequence stars this condition may be
deemed to be fulfilled. As a result of the interaction
of the star with the external potential, its energy and
angular momentum are transferred from the orbital
motion to the stellar eigenmodes, which are then
damped either through viscosity or as a consequence
of nonlinear effects.

In Zahn’s theory of dynamical tides the boundary
between the convective core and the radiative envelope
plays a major role. For an orbital period longer
than the dynamical time of the star the main interaction
between the g-modes and the tidal potential
will occur at this boundary (Zahn 1970; Goodman
and Dicksun 1998). The resonant gravitational g-modes
are assumed to be damped through moderately
rapid dissipation in a thin layer beneath the stellar
surface (Zahn 1975). Moderately rapid dissipation
implies that the characteristic rate of viscous dissipation
at the resonant frequency is much greater than
the difference between the resonant and neighboring
frequencies (Ivanov et al. 2013). This will occur
when the mode decay time due to viscosity is much
shorter than the travel time of a wave packet with
frequencies of the order of the resonant one (i.e., of
the order of the reciprocal distance between normal
mode eigenfrequencies; Ivanov et al. 2013; Goodman
and Dicksun 1998).

A huge number of papers devoted to dynamical
tides have been published in the last several years
(Bolmont and Mathis 2016; Weinberg et al. 2012;
Ivanov and Papaloizou 2004, 2007, 2010; Essick and
Weinberg 2016; Lanza and Mathis 2016; Ogilvie
2014; Papaloizou and Ivanov 2005, 2010). Tassoul
(1980) and Smeyers and Tassoul (1987) developed
an asymptotic theory of adiabatic free oscillations
for the p- and g-modes. These works generalize
the work by Zahn (1970). The rotation effects were
considered by Rocca (1987). A theory of dynamical
tides for stars with a radiative core and a convective
envelope was developed by Ivanov et al. (2013). In
contrast to Zahn’s theory, in this paper two stellar
regions contribute to the overlap integral: the convective
envelope and the radiative region near the base
of the convective envelope. Using Zahn’s formalism,
Claret and Cunha (1997) and Kushnir et al. (2016)
calculated the orbital evolution time scales as a function
of problem parameters.

In this paper the approximations used in the original
paper of Zahn (1970) are analyzed critically. The
evolution time scales of the semimajor axis are calculated
for a binary system containing a point object
with a mass equal to the solar mass and a star with
masses of one and a half and two solar masses. Below
we show that for these stars Zahn’s theory provides
reasonable agreement with our numerical calculations
only for sufficiently long orbital periods. In the
regime of moderately rapid dissipation the tidal orbital
evolution can be faster than that in Zahn’s theory by
several orders of magnitude. In addition, we show
how the quantities appearing in Zahn’s theory and
determined for polytropic stars can be calculated in
the case of stars with a realistic structure.

\section{DEFINITION OF THE BRUNT–V$\ddot {\rm A}$IS$\ddot {\rm A}$LL$\ddot {\rm A}$
FREQUENCY}

There exist three types of eigenfrequencies for a
nonrotating star without a magnetic field: the p-, f-,
and g-modes (Christensen-Dalsgaard 1998). Zahn’s
asymptotic theory of dynamical tides was considered
in the low-frequency limit for the g-modes. The properties
of the g-modes are determined by the Brunt–
V$\ddot {\rm a}$is$\ddot {\rm a}$ll$\ddot {\rm a}$ (BV) frequency. It is specified by the following relation (Christensen-Dalsgaard 1998):
\begin{eqnarray}
 N^2=g\left(\frac{1}{\Gamma p}\frac{\partial p}{\partial r}-
 \frac{1}{\rho}\frac{\partial\rho}{\partial r}\right),
\end{eqnarray}
where p is the pressure, c is the density, g is the gravitational
acceleration, r is the radial coordinate, and
$\Gamma=(\partial\ln p/\partial\ln\rho)_{\rm ad}$ is the adiabatic index. However,
this definition of the BV frequency is difficult to apply
to many stars. This is related to the numerical errors,
to the calculation of the derivative of the density, and
to the fact that there exist regions in stars where
both terms in parentheses can be of the same order
of magnitude. Therefore, a different definition of the
BV frequency is used (Brassard et al. 1991):
\begin{eqnarray}
 N^2=\frac{g^2\rho}{P}\frac{\chi_{T}}{\chi_\rho}\bigg[\nabla_{\rm ad}-\nabla-
 \sum_{i=1}^{N-1}\frac{\chi_{X_i}}{\chi_T}\frac{d \ln X_i}{d\ln p}\bigg],
\end{eqnarray}
where
\begin{eqnarray}
 \nabla_{\rm ad} = \left(\frac{\partial\ln T}{\partial\ln P}\right)_{{\rm ad},X_i};
 \chi_{T}=\left(\frac{\partial\ln p}{\partial\ln T}\right)_{\rho,X_i},\nonumber\\
 \chi_\rho = \left(\frac{\partial\ln p}{\partial\ln\rho}\right)_{T,X_i};
 \chi_{X_i}=\left(\frac{\partial\ln p}{\partial\ln X_i}\right)_{\rho,T,X_{i\ne j}},
\end{eqnarray}
the temperature gradient
\begin{eqnarray}
  \nabla=\frac{\partial\ln T}{\partial\ln P},
\end{eqnarray}
$X_i$ is the mass fraction of atoms of species i, and
\begin{eqnarray}
 \sum_{i=1}^{N-1}X_i+X_N=1.
\end{eqnarray}
It is convenient to divide the BV frequency into two
parts (Chernov 2017): the structure term
\begin{eqnarray}
 N^2_{\rm st}=\frac{g^2\rho}{P}\frac{\chi_{T}}{\chi_\rho}\bigg[\nabla_{\rm ad}-\nabla\bigg]
\end{eqnarray}
and the composition term
\begin{eqnarray}
 N^2_{\rm com}=-\frac{g^2\rho}{P}\frac{1}{\chi_\rho}
 \sum_{i=1}^{N-1}\chi_{X_i}\frac{d \ln X_i}{d\ln p}.
\end{eqnarray}
The composition term is related to the change in
stellar chemical composition due to nuclear reactions,
and it is of great importance at the boundary of the
convective and radiative regions.

When considering Zahn’s theory of dynamical
tides (Zahn 1970), we imposed significant constraints
on the BV frequency. The BV frequency $N^2$ was
assumed to have a first-order pole on the stellar
surface (see Eq. (16a) in Zahn (1970)). This behavior
of the BV frequency is typical for polytropic stars. In
stars with a realistic structure there is an atmosphere
on the stellar surface that significantly distorts the
BV frequency $N^2$. Therefore, the BV frequency on the
stellar surface actually has a more complex structure.

\begin{figure}[h]
\includegraphics[width=0.45\textwidth]{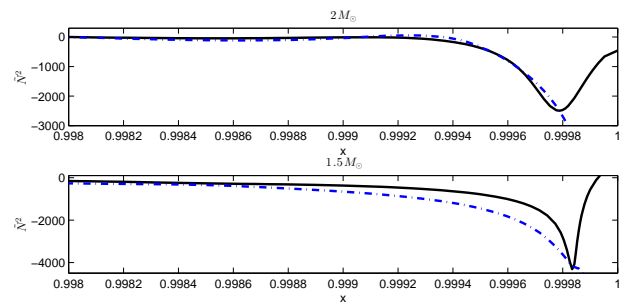}
\caption{The BV frequency near the stellar surface.}
\label{fig1}
\end{figure}

In Fig. 1 the BV frequency is plotted against the radius
near the surface for a star with a mass of one and
a half and two solar masses and an age $t = 7.82\times10^8$
and $1.24 \times 10^8$ yr, respectively (see Chernov 2017).
The solid line indicates a realistic curve of the BV
frequency for the star of two solar masses on the upper
panel and the star of one and a half solar masses on
the lower panel. The dash.dotted line indicates the
curve calculated using Eq. (16a) from Zahn (1970).
The effective polytropic index $\nu$ (Zahn 1970), which
is defined as
\begin{eqnarray}
 \frac{\nu+1}{\nu}=\lim_{x\rightarrow1}\frac{d\log P}{d\log\rho}
\end{eqnarray}
is $\nu = 2.92$ for the star of two solar masses at $x = 1$,
where $x = r/R$ is the dimensionless radial coordinate
and R is the stellar radius, and $\nu = 0.95$ for the star
of one and a half solar masses at $x = 0.999$. It can
be seen from Fig. 1 that in realistic stars the BV frequency
near the stellar surface has a sag, and Zahn’s
assumption describes this sag only partly. This assumption
works more poorly as one approaches the
stellar surface. It is worth noting that the calculations
of the effective polytropic index are also complicated
due to the presence of an atmosphere and can lead
to significant errors. For example, for the star of one
and a half solar masses the parameter $\nu$ was chosen
at radius $x = 0.999$. This approximation turned out
to describe best the BV frequency precisely for this
radius. This approximation does not work for the
parameter $\nu$ at $x = 1$. The BV frequency can also
take zero values (near the stellar surface), which was
disregarded by Zahn (1970).

\begin{figure}[h]
\includegraphics[width=0.45\textwidth]{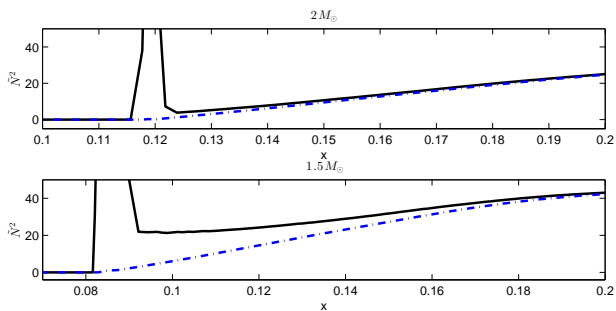}
\caption{The BV frequency near the boundary of the convective and radiative regions.}
\label{fig2}
\end{figure}

The next assumption is that the BV frequency is
zero exactly at the boundary of the convective and
radiative regions (below the subscript f in a particular
quantity $A_f$ will mean that the quantity is taken at
this boundary) and is a linear function of radius in
the radiative region near the boundary with the convective
region. In reality, only the structure term of
the BV frequency is zero exactly at the boundary of
this region; the composition term of the BV frequency
has a spike related to hydrogen burning. Thus, the
approximation used by Zahn works only for stars in
which the thermonuclear burning is sufficiently weak
and the composition term of the BV frequency is
approximately zero, $N^2_{com}\approx0$. In Fig. 2 the BV frequency
is plotted against the radius near the boundary
of the convective and radiative regions. The upper
panel is for the star of two solar masses and an age
$t = 1.24\times 10^8$ yr; the lower panel is for the star of one
and a half solar masses and an age $t = 7.82\times10^8$ yr.
The solid curve indicates the BV frequency, while
the dash-dotted curve indicates the structure term
of the BV frequency. It follows from Fig. 2 that only
the structure term becomes zero at the boundary of
the convective and radiative regions $N^2_{st} \sim(x-x_f)$,
while the BV frequency itself at the boundary of the
convective and radiative regions is not zero but has a
spike.

The approximations used by Zahn in developing
the theory of dynamical tides are not quite correct for
stars with a realistic structure. This is because the
method of solving second-order ordinary differential
equations proposed by Olver (1974), which is a mathematically
rigorous method, is used in Zahn’s theory.
The essence of Olver’s method is briefly as follows. An equation of the form
\begin{eqnarray}
 \frac{d^2w}{dz^2}=\{up(z)+q(z)\}w
\end{eqnarray}
is considered for a large value of the parameter u.
Olver (1954) considered three cases: when the function
p(z) has no singularities, when the function p(z)
has one singularity in the form of a simple zero at
some point $z_0$, and when the function p(z) has one
singularity in the form of a second-order pole. In the
latter case, the function q(z) can also have a first- or
second-order pole at the same point $z_0$. Olver (1956)
considered the fourth case where the function p(z)
has one singularity in the form of a first-order pole,
while the function q(z) has a second-order pole at
the same point. Thus, different forms of solutions
are possible, depending on which singularities the BV
frequency has. Zahn’s above assumptions about the
singularities in the BV frequency are important for
this method, and not quite correct approximations
can affect significantly the solution of our problem.

\section{OVERLAP INTEGRALS}

The overlap integrals play an important role in the
theory of dynamical tides. These are defined as follows
(Press and Teukolsky 1977):
\begin{eqnarray}
 Q=\frac{R^2\int {\rm grad}[\left(\frac{r}{R}\right)^nY_n^m]\cdot\delta \vec{r}_i \,dm}
 {\sqrt{\int (\delta \vec{r}_i\cdot\delta \vec{r}_i)dm}},
\end{eqnarray}
where $\delta\vec{r}_i$ is the Lagrangian displacement vector and
$Y_n^m$ is a spherical harmonic. The overlap integral
serves as a measure of the excitation efficiency of
normal modes in a star by tidal forces (Press and
Teukolsky 1977) and is of great importance in the
theory of dynamical tides. Below we will present our
results for the quadrupole case n = 2, because it is
this part that makes the largest contribution to the
overlap integral (Press and Teukolsky 1977). A different
quantity, the resonance coefficient, which differs
from the overlap integral by the normalization factor,
was calculated in Zahn’s theory. The overlap integral
is easy to express via the resonance coefficient. Using
Eqs. (32) and (34) from Zahn (1970), we obtain an
expression for the overlap integral in Zahn’s theory of
dynamical tides:
\begin{eqnarray}
 Q &=& \tilde{Q}\sqrt{M}R,\nonumber\\
\tilde{Q} &=& (-1)^i\frac{\sqrt{2\pi }H_n}{3^{1/6}\Gamma\left(\frac{2}{3}\right)}\sqrt{\frac{R^3}{M}\rho_f}
\left(\frac{\sigma^2R^3}{GM}\right)^{\frac{17}{12}}\times\nonumber\\
&\times&\frac{\bigg[(\frac{-R^3gA}{GMx^2})^{'}_{x_f}\bigg]^{-\frac{1}{6}}}{(n(n+1))^{\frac{11}{12}}
\sqrt{\int\limits_{x_f}^1\sqrt{-\frac{R^3gA}{GMx^2}}dx}}.
\label{Q}
\end{eqnarray}
where M is the stellar mass, R is the stellar radius,
G is the gravitational constant, and $H_n$ is a constant
coefficient determined by the convective stellar core
(see Eq. (36) in Zahn (1970)). In contrast to the resonance
coefficient, the overlap integral is a quantity
with dimensions $\sqrt{M}R$. Thus, a universal frequency
dependence of the overlap integral $Q\sim\omega^{17/6}$, where
the exponent does not depend on the stellar structure,
is obtained in Zahn's theory. The proportionality
factor of the overlap integral is determined entirely
by the stellar density and the BV frequency at the
boundary of the convective and radiative regions. In
contrast to the resonance coefficient, the overlap integral
does not depend on the stellar density and the
BV frequency at the stellar surface. The evolution of
the overlap integral is related to the change in stellar
density and BV frequency. This is responsible for the
change of the overlap integral with time.

Difficulties in calculating the contribution from the
convective core to the overlap integral also arise in
Zahn’s theory (Kushnir et al. 2016). The difficulties
stem from the fact that the prefactor consists of two
terms that depend strongly on stellar parameters. The
first and second terms take fairly large and fairly small
values, respectively. As a result of their multiplication,
the error can be compensated (Kushnir et al. 2016).
Kushnir et al. (2016) showed that the prefactor could
be calculated quite accurately from a polytrope stellar
model, and an alternative form of the angular momentum
transfer through the quantities that are determined
at the boundary of the convective and radiative
regions was derived.

\section{COMPARISON OF ZAHN’s THEORY WITH
OUR NUMERICAL CALCULATIONS}

In this section we will compare the results of
Zahn’s analytical theory of dynamical tides (11)
with our numerical calculations for a realistic stellar
structure. We considered stars of one and a half and
two solar masses. We chose stellar models similar
to the stellar models 2b, 2c, 2d and 1.5b, 1.5c from
Chernov et al. (2013). The table presents the stellar
parameters and the quantities that were calculated for
each stellar model using Zahn’s theory. To calculate
the integral in the fourth row of the table, we used
the full BV frequency. To calculate the derivative
at point $x_f$, we used only the structure term of the
BV frequency, because this term behaves smoothly
at this point. All these stars have a convective core
and a radiative envelope. The stars were modeled
with the MESA software package (Paxton 2011,
2013, 2015); the data obtained were then interpolated
by the Steffen (1990) method to two million points.
Using these data, we solved the eigenfrequency and
eigenfunction problem and calculated the overlap integrals
in the adiabatic approximation. The methods
of finding the eigenfrequencies and eigenfunctions
are described in Christensen-Dalsgaard (1998). The
fourth-order Runge–Kutta method was used to solve
the differential equations.
\begin{figure}
\includegraphics[width=0.45\textwidth]{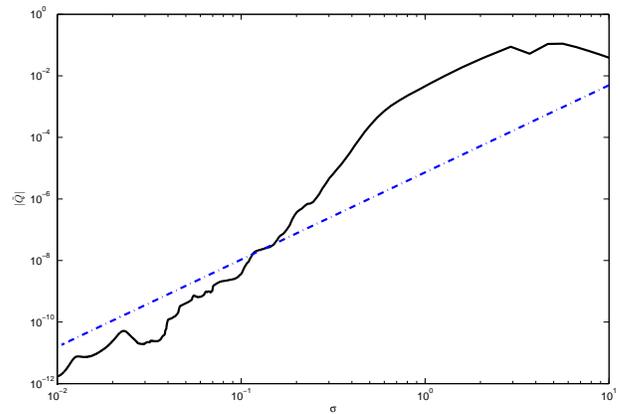}
\caption{Overlap integral versus frequency for model 2b.}
\label{figQ2i2}
\end{figure}
\begin{figure}
\includegraphics[width=0.45\textwidth]{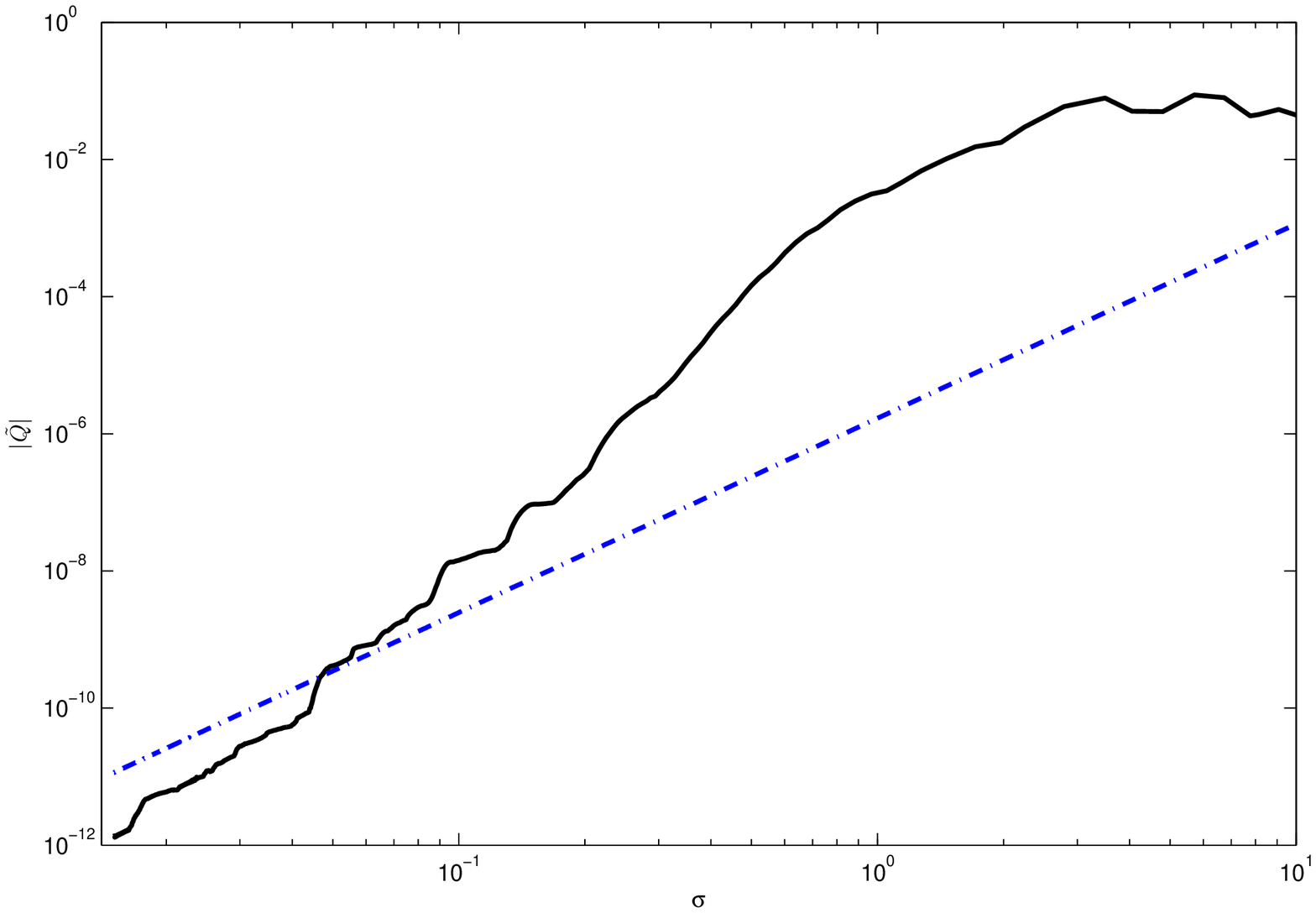}
\caption{Overlap integral versus frequency for model 2c.}
\label{figQ2i3}
\end{figure}
\begin{figure}
\includegraphics[width=0.45\textwidth]{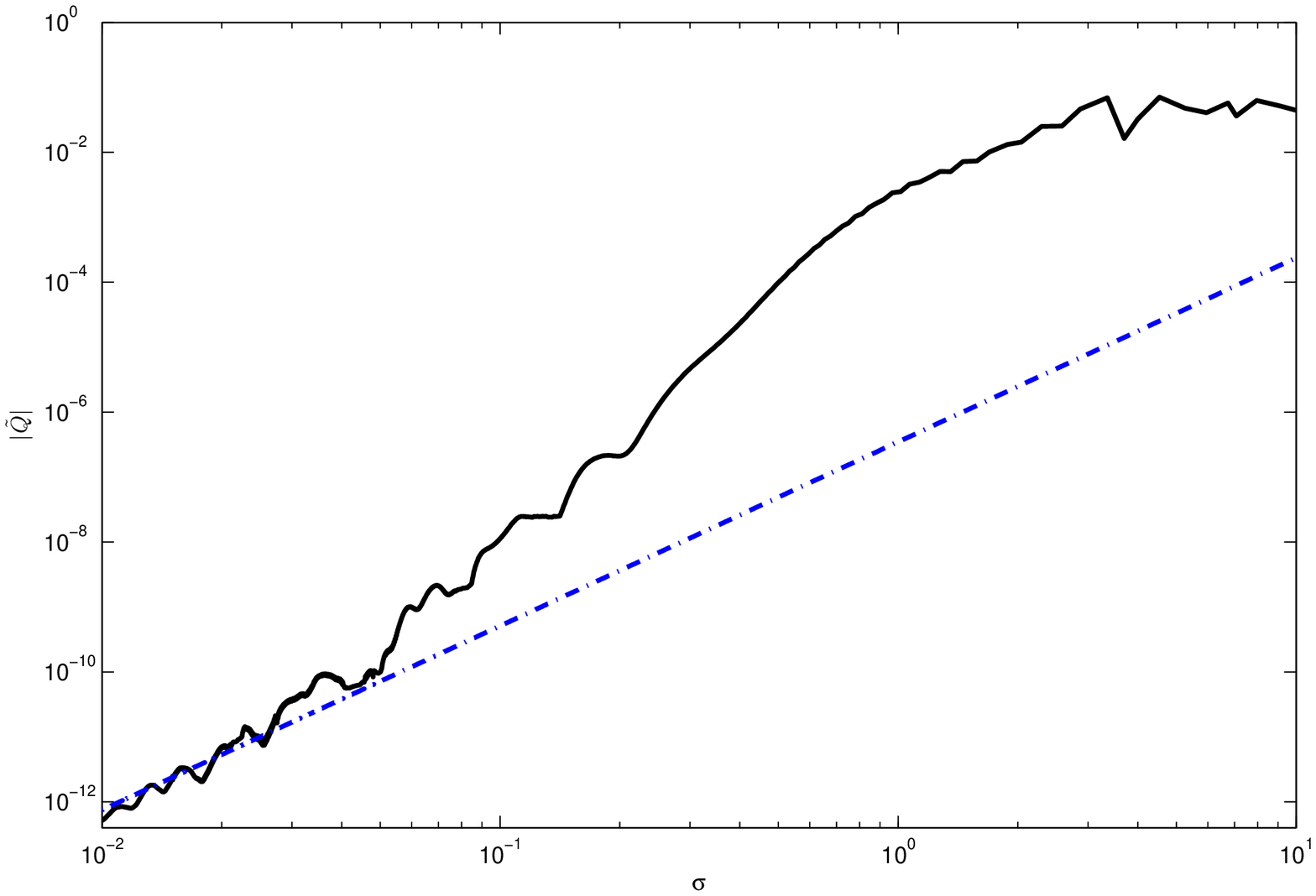}
\caption{Overlap integral versus frequency for model 2d.}
\label{figQ2i4}
\end{figure}
\begin{figure}
\includegraphics[width=0.45\textwidth]{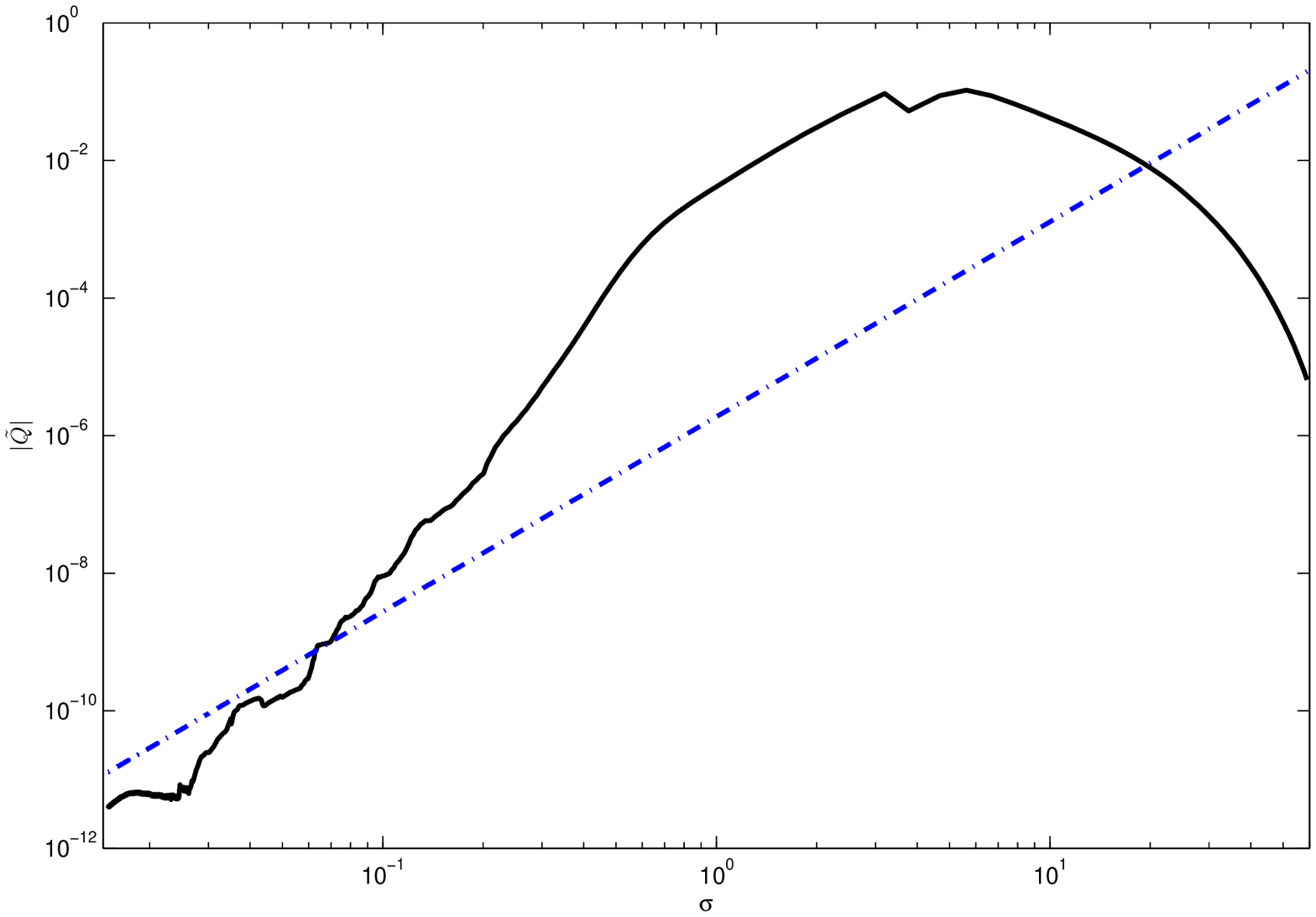}
\caption{Overlap integral versus frequency for model 1.5b.}
\label{figQ690}
\end{figure}
\begin{figure}
\includegraphics[width=0.45\textwidth]{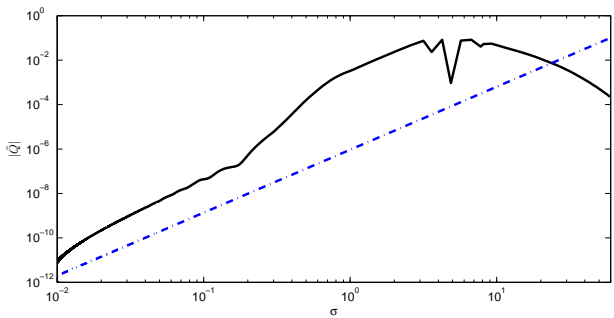}
\caption{Overlap integral versus frequency for model 1.5c.}
\label{figQ350}
\end{figure}

The overlap integrals are shown in Figs. 3–7.
The solid line indicates our numerical calculations;
the dash–dotted line indicates the analytical curve
obtained from Zahn’s formula (11). There is a
close asymptotic correspondence of the theory to the
numerical calculations only for model 2d (Fig. 5a).
There is a discrepancy by one order of magnitude
or more for other models. This discrepancy may be
related to the rough approximations that were used in
Zahn’s theory to describe the BV frequency.

For all values of the overlap integral there is a
successive change of sign in the dependence on the
number of nodes i of the eigenfunctions, which is
consistent with Eq. (11) (the factor $(-1)^i$), except
for some cases. At frequencies $\omega\sim0.4$ there is a
glitch, when the overlap integral does not change
its sign for two neighboring frequencies. This may
stem from the fact that the two solutions obtained
near the stellar surface and at the boundary of the
convective and radiative regions are joined in the
theory when finding the eigenfrequencies of stellar
oscillations in the radiative region. However, since
the joining occurs at moderately low frequencies, the
next orders of smallness in frequency can contribute
to the oscillation phase. As a result, the prefactor
of the overlap integral decreases sharply, which is
reflected in the sharp oscillations in Figs. 3-6. At
lower frequencies the contribution to the oscillation
phase will decrease, and the overlap integrals are
asymptotically smoothed out.

Some structure in the from of sags can also be
seen in Figs. 4-7 for the overlap integral at frequencies
$\sigma\sim4$. This may be related to the interaction of
the g- and p-modes (Chernov et al. 2013).

In Figs. 8 and 9 the evolution time scales of the
semimajor axis are plotted against the orbital period
of a point object around a star of one and a half
and two solar masses, respectively (see Eq. (131)
from Ivanov et al. (2013)). The stars are assumed
to be nonrotating ones; the mass of the point object
is equal to the solar mass. The orbit of the
point object is assumed to be nearly circular; the
eccentricity is approximately zero, $e\rightarrow0$. The solid
line indicates our numerical calculations; the dashed
line indicates our calculations using Zahn's overlap
integral (11). For an orbital period of the order of one
day ($P_{orb}\approx1$) the orbital evolution time derived from
Zahn's formula is longer than that in our numerical
calculations by several orders of magnitude. This is
easy to explain, because for an orbital period of one
day the characteristic eigenfrequencies lie within the
range $\omega\sim0.3-0.8$. Zahn's asymptotic theory at such
frequencies gives overlap integrals than are smaller
than those in our numerical calculations by several
orders of magnitude.
\begin{figure}
\includegraphics[width=0.45\textwidth]{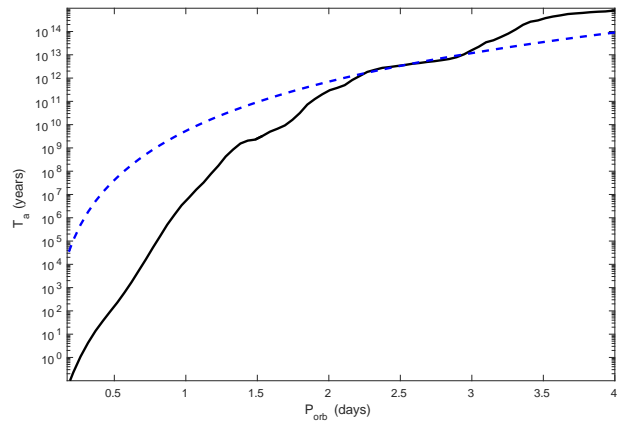}
\caption{Evolution time scale of the semimajor axis for model m2b.}
\label{Tam2i2}
\end{figure}
\begin{figure}
\includegraphics[width=0.45\textwidth]{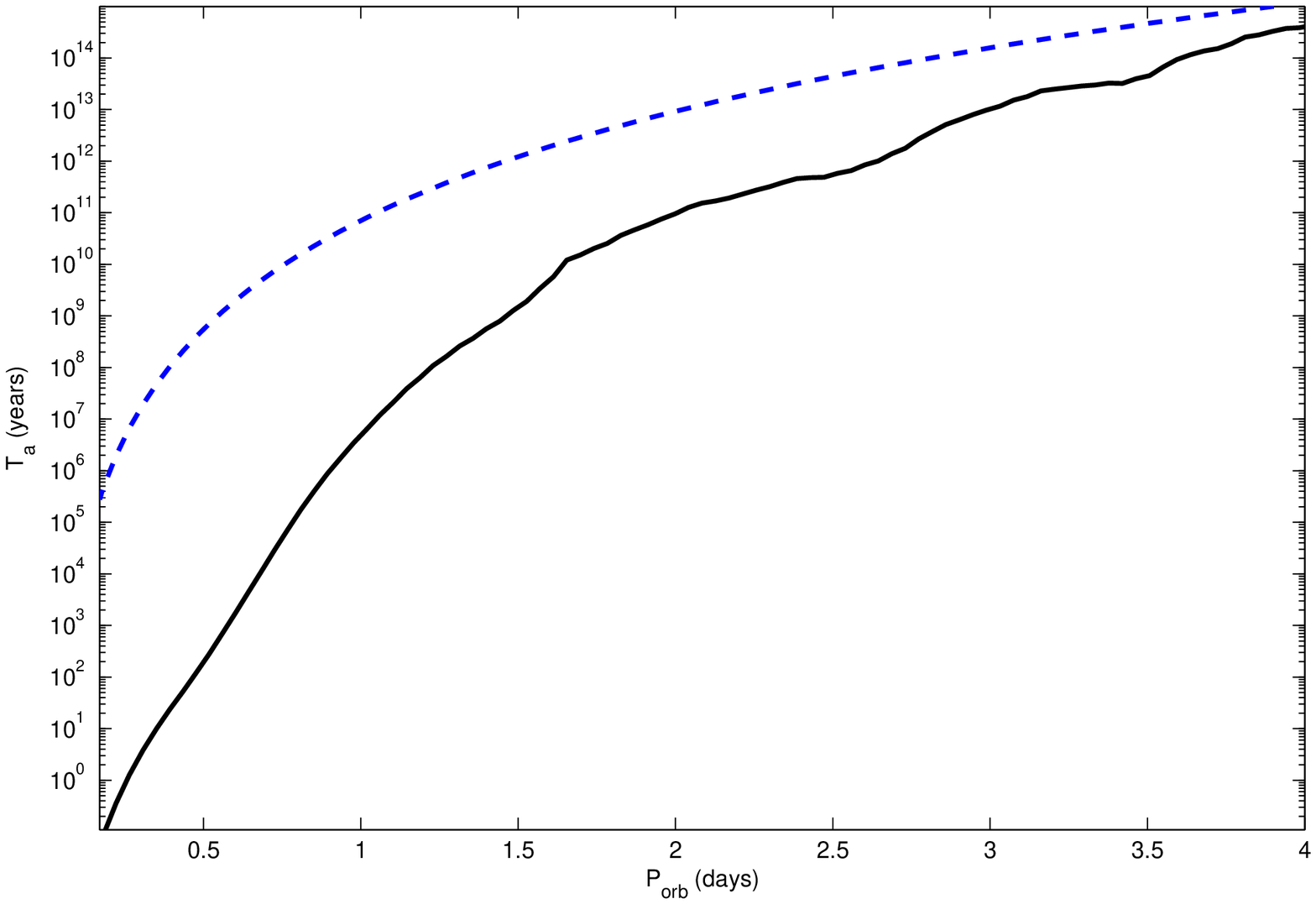}
\caption{Evolution time scale of the semimajor axis for model m1.5b.}
\label{Tam1.5x690}
\end{figure}

\begin{table}
\caption{Stellar parameters and characteristic quantities from Zahn’s theory calculated for our models. The mass and radius are measured in solar masses and radii, respectively; the age is in years.}
\label{tab}
\begin{tabular}{c|c|c|c|c|c}
\hline
 & 2b & 2c & 2d & 1.5b & 1.5c \\
 \hline
 Mass & 2 & 2 & 2 & 1.5 & 1.5\\
 \hline
 Radius & 1.63 & 2.25 & 2.91 & 1.46 & 1.82 \\
 \hline
 Age & 2.93e7 & 5.93e8 & 8.44e8 & 5.96e7 & 1.58e9 \\
 \hline
 $\sqrt{\int\limits_{x_f}^1\!\!\!\sqrt{\!\frac{\tilde{N}^2}{x^2}}\,dx}$ & 2.87 &
 3.81 & 5.07 & 3.12 & 3.94 \\
 \hline
 $\frac{R^3}{M}\rho_f$ & 16.27 & 42.93 & 117.1 & 23.15 & 49.95 \\
 \hline
 $\left(\frac{\tilde{N}^2}{x^2}\right)^{'}_{|x_f}$ & 1.5e4 & 5e4 & 1e5
 & 1.8e4 & 1.1e5 \\
 \hline
 $H_2$ & 8.63e-5 & 1.99e-5 & 3.73e-6 & 2.09e-5 & 1.20e-5\\
 \hline
\end{tabular}
\end{table}

\section{CONCLUSIONS}

We showed that for the models of stars with a
convective core and a radiative envelope with masses
of one and a half and two solar masses in the approximation
of moderately rapid dissipation and a dense
spectrum, the tidal evolution is much faster than
follows from Zahn’s theory for an orbital period of
the order of one day. Such significant discrepancies
probably stem from the fact that in the models for the
tidally excited resonant terms under consideration,
the asymptotic expression for the overlap integrals
derived by Zahn is smaller than the numerical one by
several orders of magnitude (see Figs. 3–7).

The fact that our approach gives much shorter
tidal evolution time scales of the semimajor axis may
turn out to be very important for questions related to
the tidal evolution of both binary stars and systems of
exoplanets, because in both cases the existing theoretical
approaches constructed “from first principles”
often give time scales that are too long to explain the
observations. Note, however, that since the resonant
frequencies and, accordingly, the distances between
them are relatively large in our case, the question
arises as to whether the approximation of moderately
rapid dissipation is valid.

However, despite all its shortcomings, Zahn’s theory
of dynamical tides is one of the few theories that
qualitatively describes the dynamical tides in stars
with a convective core and a radiative envelope and
explains many observations. Therefore, this theory
deserves rapt attention, but it should be applied with
caution to specific stars with a realistic structure.

\section{ACKNOWLEDGMENTS}

I am grateful to P.B. Ivanov and R.R. Rafikov,
who read the paper and made a number of valuable
remarks, and to J. Papaloizou for a fruitful
discussion. This work was financially supported by
the Russian Foundation for Basic Research (project
nos. 140200831, 150208476, 160201043), grant
NSh-6595.2016.2 from the President of the Russian
Federation for State Support of Leading Scientific
Schools, and Program 7 of the Presidium of the
Russian Academy of Sciences.


\begin{thebibliography}{99}

\bibitem{Bolmont} E. Bolmont, S. Mathis, Celestial Mechanics. and Dynamical Astronomy 126, 275 (2016).

\bibitem{Brassard} P. Brassard, G. Fontaine, F. Wesemael, S.D. Kawaler, M. Tassoul, Astrophys. J. 367, 601 (1991).

\bibitem{Chernov} S.V. Chernov, Astron. Let. 43(3), 186 (2017).

\bibitem{ChernovPapIvanov} S.V. Chernov, J.C.B. Papaloizou, P.B. Ivanov, MNRAS 434, 1079 (2013).

\bibitem{Dalsgaard} J. Christensen-Dalsgaard, Lecture notes on stellar oscillations, 4th edn. (1998).

\bibitem{Claret} A. Claret, N.C.S. Cunha, Astron.Astrophys. 318, 187 (1997).

\bibitem{Cowling} T.G. Cowling, MNRAS 101, 367 (1941).

\bibitem{Essick} R. Essick, N. Weinberg, Astrophys.J. 816, 21 (2016).

\bibitem{GoodmanDicksun} J. Goodman, E. Dicksun, Astrophys.J. 507, 938 (1998).

\bibitem{IvanovPapaloizou2004a} P.B. Ivanov, J.C.B. Papaloizou, MNRAS 347, 437 (2004).

\bibitem{IvanovPapaloizou2004b} P.B. Ivanov, J.C.B. Papaloizou, MNRAS 353, 1161 (2004).

\bibitem{IvanovPapaloizou2007} P.B. Ivanov, J.C.B. Papaloizou, MNRAS 376, 682 (2007).

\bibitem{IvanovPapaloizou2010} P.B. Ivanov, J.C.B. Papaloizou, MNRAS 407, 1609 (2010).

\bibitem{IvanovPapChernov} P.B. Ivanov, J.C.B. Papaloizou, S.V. Chernov, MNRAS 432, 2339 (2013).

\bibitem{Kushnir} D. Kushnir, M. Zaldarriaga, J. Kollmeier, R. Waldman, ArXiv:1605.03810v1.

\bibitem{Lanza} A.F. Lanza, S. Mathis, Celestial Mechanics. and Dynamical Astronomy 126, 249 (2016).

\bibitem{Ogilvie} G. Ogilvie, Annual Rev. Astron. Astrophys. 52, 171 (2014).

\bibitem{Olver1954} F.W.J. Olver, Phil. Trans. Royal Soc. London Ser. A 247, 307 (1954).

\bibitem{Olver1956} F.W.J. Olver, Phil. Trans. Royal Soc. London Ser. A 249, 65 (1956).

\bibitem{OlverBook} F.W.J. Olver, Asymptotics and special functions, Academic Press, (1974).

\bibitem{PapaloizouIvanov2005} J.C.B. Papaloizou, P.B. Ivanov, MNRAS 364, L66 (2005).

\bibitem{PapaloizouIvanov2010} J.C.B. Papaloizou, P.B. Ivanov, MNRAS 407, 1631 (2010).

\bibitem{Paxton2011} B. Paxton et al., Astrophys. J. Supp. Ser. 192, 3
(2011).

\bibitem{Paxton2013} B. Paxton et al., Astrophys. J. Supp. Ser. 208, 4
(2013).

\bibitem{Paxton2015} B. Paxton et al., Astrophys. J. Supp. Ser. 220, 15
(2015).

\bibitem{PressTeukolsky} W.H. Press, S.A. Teukolsky, Astrophys.J. 213, 183 (1977).

\bibitem{Rocca} A. Rocca, Astron. Astrophys. 175, 81 (1987).

\bibitem{SmeyersTassoul} P. Smeyers, M. Tassoul, Astrophys.J. Suppl. 65, 429 (1987).

\bibitem{Steffen1990} M. Steffen, Astron. Astrophys. 239, 443 (1990).

\bibitem{Tassoul1980} M. Tassoul, Astrophys. J. Suppl. 43, 469 (1980).

\bibitem{Weinberg} N. Weinberg, P. Arras, E. Quataert, J. Burkart, Astrophys.J. 751, 37 (2012).

\bibitem{Zahn1970} Zahn, Astron. Astrophys. 4, 452 (1970).

\bibitem{Zahn1975} Zahn, Astron. Astrophys. 41, 329 (1975).

\bibitem{Zahn1977} Zahn, Astron. Astrophys. 57, 383 (1977).


\end{thebibliography}
\end{document}